\documentclass[preprint,aps,nofootinbib,tightenlines,
    byrevtex]{revtex4}
\usepackage{graphicx}%
\usepackage{dcolumn}
\usepackage{amsmath}
\usepackage{epsfig}
\usepackage{latexsym}
\usepackage{amssymb}
\usepackage{mathrsfs}

\newcommand{\3}{\ss}

\def\pislash{ {\pi\hskip-0.6em /} }

\def\nopi{ {\rm EFT}(\pislash) }

\def\nrcpt{NR\raise.4ex\hbox{$\chi$}PT\ }
\def\ket#1{\vert#1\rangle}
\def\bra#1{\langle#1\vert}
\def\ltap{\ \raise.3ex\hbox{$<$\kern-.75em\lower1ex\hbox{$\sim$}}\ }
\def\gtap{\ \raise.3ex\hbox{$>$\kern-.75em\lower1ex\hbox{$\sim$}}\ }

\def\frac#1#2{{\textstyle{#1\over#2}}}
\def\darr#1{\raise1.5ex\hbox{$\leftrightarrow$}\mkern-16.5mu #1}
\def\){\right)}
\def\({\left( }
\def\]{\right] }
\def\[{\left[ }
\def\si{{}^1\kern-.14em S_0}
\def\siii{{}^3\kern-.14em S_1}
\def\diii{{}^3\kern-.14em D_1}

\newcommand{\LambdaNoPion}{\ensuremath{\Lambda_{\pi\hskip-0.4em /}}}

\newcommand{\calH}{\mathcal{H}}

\newcommand{\dis}{\displaystyle}
\newcommand{\ii}{\mathrm{i}}
\newcommand{\dd}{\mathrm{d}}

\begin{document}
\bibliographystyle{plain}
\def\nopi{ {\rm EFT}(\pislash) }

\title{Triton Electric Form Factor at Low-Energies}

\author{H. Sadeghi}\email{H-Sadeghi@araku.ac.ir}
 \affiliation{Department of Physics, University of Arak, P.O.Box 38156-879, Arak,
 Iran.}

\vspace{4cm}

\begin{abstract}
\vspace{0cm}

 Making use of the Effective Field Theory(EFT) expansion recently
developed by the authors, we compute the charge form factor of
triton up to next-to-next-to-leading order (N$^2$LO). The
three-nucleon forces(3NF) is required for renormalization of the
three-nucleon system and it effects are predicted for process and is
qualitatively supported by available experimental data. We also show
that, by including higher order corrections, the calculated charge
form factor and charge radius of $^3$H are in satisfactory agreement
with the experimental data and the realistic Argonne $v_{18}$
two-nucleon and Urbana IX potential models calculations. This method
makes possible a high precision few-body calculations in nuclear
physics. Our result converges order by order in low energy expansion
and also cut-off independent.

\begin{tabular}{c}
\end{tabular}

\begin{tabular}{rl}

keywords:&\begin{minipage}[t]{11cm} effective field theory,
three-body system, three-body force, triton form factor
  \end{minipage}

\end{tabular}
\end{abstract}

\vskip 1.0cm \noindent

\maketitle

\section{Introduction}

The few-body problem has now reached the stage when the main
interest is centered not on theoretical techniques but on the
application of these techniques to problems in nuclear, particle and
atomic physics. Among these theories, EFT is a powerful tool to
calculate low-energy observables in a systematic way. Pionless EFT
has been applied to two- and three-nucleon systems
~\cite{kaplan,Chen,Beane,3stooges_boson,3stooges_doublet,4stooges,chickenpaper,platter-FBS40,20},
during the last few years. It is ideally suited to exploit a
separation between scales and can use with contact interactions at
low-energies. Some of these low-energy observables such as
neuron-deuteron radiative capture and the two-body
photodisintegration of the triton have been calculated using
pionless EFT and insertion of the three-body force, up to
N$^2$LO~\cite{Sadeghi1,Sadeghi2,Sadeghi3,Sadeghi4}. In these
calculation we find a simple way for inclusion of external currents.
The evaluated cross sections have been compared with experimental
data and other modern realistic two- and three-nucleon forces
AV18/UrbIX potential models calculations. We are also going to study
the structural functions of three-body systems in future
calculations. Among them, the study of the three-body nuclear
physics involving the charge form factors and charge radius of $^3$H
is calculated by using pionless EFT. It have been investigated in
theoretical and experimental works over the past decays.

The effect of meson exchange currents on the charge and magnetic
form factors of $^3$H was investigated by Maize et~al.~\cite{Maize}
and ref there in. Inclusion of meson exchange currents considerably
improves the impulse approximation fits to the experimental data. An
other calculation reported in  include single-$\Delta$ isobar
admixtures in the three-nucleon wave function~\cite{Strueve}.

Recently, a very good agreement reached with the calculation of the
charge form factor of $^3$H~\cite{Marcucci,Golak}. They calculated
the electromagnetic form factors of the three-nucleons $^3$H with
wave functions obtained with the Argonne v18 two-nucleon and Urbana
IX three-nucleon interactions. The two-body currents required by
current conservation with the v18 interaction as well as those
associated with N$\Delta$ transition currents and the currents of
$\Delta$ resonance components in the wave functions, are considered
for these calculations. In this way, explicit three-nucleon current
operators associated with the two-pion exchange three-nucleon
interaction arising from irreducible S-wave pion-nucleon scattering
is constructed and it is shown to have very little effect on the
calculated magnetic form factors.

More recent calculation of the triton charge form factor at leading
order in $\ell/|a|$ and extract the charge radius has been performed
by Platter et~al.~\cite{Platter}. They studied the correlation
between the triton binding energy and charge radius and relate it to
the three-body parameter $L_3$ that parameterizes the Phillips and
Tjon lines in effective theory.

In experimental works, the $^3$H charge form factor was measured in
the range 0.29  $< q^2 <$  1 fm$^{-2}$ by Beck et~al.~\cite{Beck1}.
After few years, the charge and magnetic form factors were also
measured in the range 0.0477 $< q^2 <$ 2.96 fm$^{-2}$ ~\cite{Beck2}.
The experimental result of triton charge radius is quoted as $r_c$=
1.63 $\pm$ 0.03 fm~\cite{Beck2}; as $r_c$ = 1.76 $\pm$ 0.04
fm~\cite{Juster};  as $r_c$ = 1.81 $\pm$ 0.05 fm~\cite{Martino}; and
as $r_c$ = 1.755 $\pm$ 0.086 fm~\cite{Amroun}.

In this paper, we study the charge form factor and chargr radius of
the triton to next-to-next-to-leading order (N$^2$LO). Our results
show that, by including higher order corrections, the calculated
charge form factor of $^3$H is in satisfactory agreement with the
modern nucleon-nucleon potential AV18 together with the three-
nucleon force UrbanaIX. This  result has several implications. The
first and most important one is the remarkable success of EFT
improving for simple inclusion of external currents. The second is
the indication that the model makes possible a high precision
three-body calculations in nuclear physics. It now appears to
explain the structural functions of three-body systems. The third is
remarkable quality of the EFT low-energiy calculations. The result
converges order by order in low energy expansion, model independent
and also cut-off independent.  It should be considered that it is a
quantitative question based on current choices and gauge invariance
of nuclear force models to reveal signatures by switching on and off
3N forces.

This paper is divided into five main sections. Section II contains a
brief description of the calculation of the the bound three-body
systems.  Faddeev integral equation, Lagrangian and kernel for the
reaction involved in our consideration and three-body forces are
given in this section. In Section III, we calculate the charge form
factor of the triton by using EFT. Calculation of the charge form
factors, the relevant diagrams, and parameters are discussed in this
section. In section IV, comparisons of our results with the
corresponding experimental and theoretical data are given. Finally
section V contains a concluding discussion.
\begin{figure}[!htp]
\begin{center}
\includegraphics*[width=.8\textwidth]{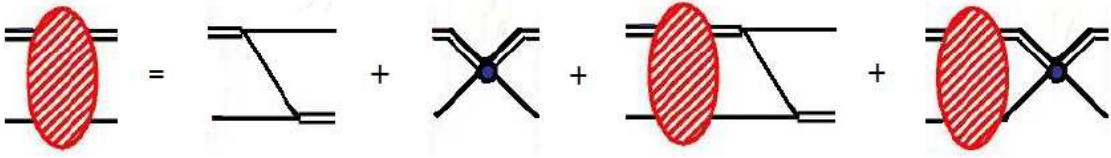}
\caption{Faddeev equation for the three-body amplitude including the
three-body force. Line shows nucleon and thick solid line is
propagator of the two intermediate auxiliary fields $d_s$ and $d_t$.
The three-body interactions with strength $\mathcal{H}$, denoted by
the dot.}
\end{center}
\end{figure}

\section{Brief review of the bound three-body system}
\label{section:Triton Charge Form factor}

The derivation of the integral equation describing neutron-deuteron
scattering has been widely discussed~\cite{3stooges_doublet,20}. We
present here only the result, including the new term generated by
the two-derivative three-body force. The three-nucleon lagrangian is
given by~\cite{4stooges}:

\begin{eqnarray}\label{triton_deuteron_lag}
  {\mathcal L}&=&N^\dagger\left(\ii\partial_0+\frac{\nabla^2}{2M}\right)N
  +d_s^{A\dagger}\left(-\ii\partial_0-\frac{\nabla^2}{4M}+\Delta_s\right)d_s^A
  +d_t^{i\dagger}\left(-\ii\partial_0-\frac{\nabla^2}{4M}+\Delta_t\right)
  d_t^i\\ \nonumber
  &&+t^\dagger\left(\ii\partial_0+\frac{\nabla^2}{6M}+\frac{\gamma^2}{M}+
    \Omega\right)t -g_s\left( d_s^{A \dagger} (N^T P^A N) +\text{H.c.}
  \right) -g_t\left( d_t^{i \dagger} (N^T P^i N) +\text{H.c.}
  \right)\\
  &&-\omega_s \left( t^\dagger (\tau^A N) d_s^{A} +\text{H.c.} \right)
  -\omega_t \left( t^\dagger (\sigma^iN) d_t^{i} +\text{H.c.} \right)
  +\dots\;\;, \nonumber
\end{eqnarray}
where $N$ is the nucleon iso-doublet and the auxiliary fields $t$,
$d_s^A$ and $d_t^i$ carry the quantum numbers of the $^3$H spin and
isospin doublet, $^1S_0$ di-nucleon and the deuteron, respectively.
The projectors are $P^i=\frac{1}{\sqrt{8}}\tau_2 \sigma_2\sigma^i$
and $P^A=\frac{1}{\sqrt{8}}\sigma_2 \tau_2\tau^A$, where $A=1,2,3$
and $i=1,2,3$ are iso-triplet and vector indices and $\tau^A$
($\sigma^i$) are isospin (spin) Pauli matrices.

For calculation in this channel, two amplitudes get mixed(see
fig.~1): $t_s$ describes the $d_t + N\rightarrow d_s + N$ process,
and $t_t$ describes the $d_t + N\rightarrow d_t + N$ process:

\begin{eqnarray}
 t_s(p,k)& =  \frac{1}{4}\left[3\mathcal{K}(p,k)
+2\mathcal{H}(E,\Lambda)\right]+\dis\frac{1}{2\pi}
 \int\limits_0^\Lambda \dd q\; q^2
    & \left[\mathcal{D}_s(q)\left[\mathcal{K}(p,q)+2\mathcal{H}(E,\Lambda)
      \right]
t_s(q)\right.\nonumber\\
       &&\left.+\mathcal{D}_t(q)\left[3\mathcal{K}(p,q)+2\mathcal{H}(E,\Lambda)
       \right]
t_t(q)\right] \label{int_equation_triton}\nonumber\\
 t_t(p,k)& = \frac{1}{4}\left[\mathcal{K}(p,k)
+2\mathcal{H}(E,\Lambda)\right]+\dis\frac{1}{2\pi}
 \int\limits_0^\Lambda \dd q\; q^2
  &\left[ \mathcal{D}_t(q)\left[
\mathcal{K}(p,q)+2\mathcal{H}(E,\Lambda)\right]
t_t(q)\right.\nonumber\\
       & & \left.+\mathcal{D}_s(q)
       \left[3\mathcal{K}(p,q)+2\mathcal{H}(E,\Lambda)\right]
t_s(q)\right]\;\;,
\end{eqnarray}
with
\begin{eqnarray}
  \label{eq:transvestite_prop}
\mathcal{D}_s(q)&=&
  \frac{1}{-\gamma_s+\sqrt{\frac{3}{4}(q^2-k^2)+\gamma_t^2}}+\\\nonumber
  &&+
    \frac{3r_s}{8}\frac{q^2-k^2+\frac{4}{3}\gamma_t^2}{
      \left(-\gamma_s+\sqrt{\frac{3}{4}(q^2-k^2)+\gamma_t^2}\right)^2}
    +\left(\frac{3 r_s}{8}\right)^2\frac{(q^2-k^2+\frac{4}{3}\gamma_t^2)^2}{
      \left(-\gamma_s+\sqrt{\frac{3}{4}(q^2-k^2)+\gamma_t^2
        }\right)^3}\;\;.
\end{eqnarray}
where $\mathcal{D}_{s,t}(q)=\mathcal{D}_{s,t}(E-\frac{q^2}{2M},q)$
are the propagators of deuteron.

For the spin-triplet $\mathrm{S}$-wave channel, one replaces the two
boson binding momentum $\gamma$ and effective range $\rho$ by the
deuteron binding momentum $\gamma_t=45.7025\;\mathrm{MeV}$ and
effective range $\rho_t=1.764\;\mathrm{fm}$. Because there is no
real bound state in the spin singlet channel of the two-nucleon
system, its free parameters are better determined by the scattering
length $a_s=1/\gamma_s=-23.714\;\mathrm{fm}$ and the effective range
$r_s=2.73\;\mathrm{fm}$ at zero momentum, The neutron-deuteron
$J=1/2$ phase shifts $\delta$ is determined by the on-shell
amplitude $t_t(k,k)$, multiplied with the wave function
renormalization~\cite{4stooges}.

\begin{equation}
T(k)=Z t_t(k,k).
\end{equation}
In that case, a unique solution exists in the $^2S_{1/2}$-channel
for each $\Lambda$ and vanishing three-body force, but no unique
limit as $\Lambda\to\infty$.  As long distance, phenomena must
however be insensitive to details of the short-distance physics (and
in particular of the regulator chosen), Bedaque et
al.~\cite{4stooges,20} showed that the system must be stabilized by
introducing of a three-body force in a form:
\begin{equation}
  \label{eq:calH}
   \calH(E;\Lambda)=
   \frac{2}{\Lambda^2}\sum\limits_{n=0}^\infty\;H_{2n}(\Lambda)\;
   \left(\frac{ME+\gamma_t^2}{\Lambda^2}\right)^n.
\end{equation}
which absorbs all dependence on the cutoff as
$\Lambda\to\infty$~\cite{4stooges}. At leading order,
$\gamma\ll\Lambda$, the essential observation are of order
$\mathcal{O}(Q/\Lambda)$ and are independent of energy (or momentum)
and hence can be made to vanish by only $H_0$ and not any of the
higher derivative three-body forces. At NLO, there is only a
perturbative change from the LO asymptotic of order
$\mathcal{O}(1/\Lambda^2)$(see ~\cite{4stooges}).

For N$^2$LO order, a new three-body force seems to be required. The
terms will be proportional to $ME$ arising from expanding the kernel
in powers of $q/\Lambda$ and $\sqrt{ME}/\Lambda$ and a three-body
force term which contains a dependence on the external momenta $k^2$
and $ME$ can absorb them. It means we need $H_2$ for calculations up
to N$^2$LO. It has been shown that the three-body system at N$^2$LO
can be renormalized without the need for an energy dependent
three-body force at this order~\cite{platter-FBS40}.

In the presence of an electromagnetic interactions, after performing
minimal substitution, requires that $\nabla$ operators be replaced
by covariant derivatives ${\bf D}=\nabla - i e {\bf A}$, and
time-derivatives $\partial_0$ be replaced by $D_0=\partial_0+i e
A_0$ in lagrangian eq.~(1), where ${\bf A}$ is the vector potential.

This simple perturbative expansion of the scattering amplitude is
reproduced order by order in the pionless expansion. Relativistic
corrections are encountered at N$^2$LO and effective range expansion
provides a complete description of scattering in the low-energy
region\cite{4stooges}.

\begin{figure}[!htp]
\begin{center}
 \includegraphics*[width=0.8\textwidth]{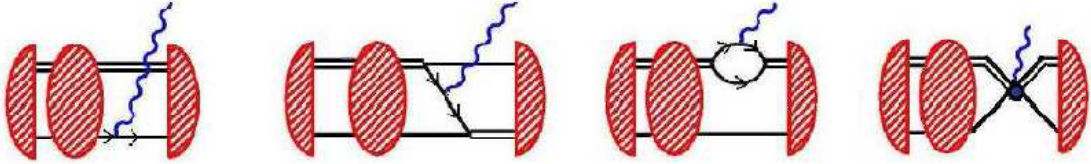}
 \caption{Diagrams that contribute to the triton charge form factor.
Wavy line shows photon. Remaining notation as in fig.~1.}
 \vspace*{-0pt}
\label{fig2}
\end{center}
\end{figure}
\section{The Charge form factor of Triton}

The electric charge form factor of the triton is measured over a
wide range of momentum transfers. The modern nucleon-nucleon
potential AV18 together with the three-nucleon force UrbanaIX
potential models and effective theory reproduce the data very well
in the kinematic regions where they are applicable. (For a recent
review, see \cite{Golak, Platter}.). In present work, we calculate
the charge form factor in the range 0 $< q^2 <$ 0.3 fm$^{-2}$ in
corresponding to recent effective theory calculation~\cite{Platter}.

A convenient expression to calculate the charge form factors of a
J=1/2 nucleus, such as the triton, is obtained by the matrix element
of the electric current, is given by:
\begin{eqnarray}
\bra{\Psi_+} J_{e}\ket{\Psi_+} &=& e \ F_C(q^2) \ \ \ ,
\label{eq:emmatdef}
\end{eqnarray}
where ${\bf q}$ is the momentum transfer and $\Psi_+$ is the
normalized ground state wave function. The dimensionless charge form
factor defined in eq.~(\ref{eq:emmatdef}) are normalized such that $
F_C(0) = 1$. The slope of the charge form factor at low-momentum
transfer defines the triton charge radius:

\begin{equation} \label{eq:chrad}
F_C({\bf q^2})= 1 - {\bf q^2} \langle r^2 \rangle/6 +\ldots\,.
\end{equation}
where $r_C\equiv \langle r^2 \rangle^{1/2}$ is the charge radius.

In the pionless EFT, the charge form factor $F_C (q^2)$ has an
expansion in powers of $Q$, $F_C (q^2) =F_C^{(0)} (q^2) +F_C^{(1)}
(q^2)+F_C^{(2)} (q^2)+...$, where the superscript denotes the order
of the contribution.

The diagrams that contribute to the deuteron charge form factor in
pionless EFT are shown in fig.~2. In addition to diagrams where the
photon couples to the nucleon, there are also couplings to the
dibarion field obtained by gauging the Lagrange density in eq.~(1).

The LO order calculation is corresponding to the leading term in a
effective range expansion and is in corresponding with the charge
form factor computed in ref.~\cite{Platter}.  At higher orders there
are contributions from higher dimension operators involving more
derivatives on the nucleon field, such as the nucleon charge radius
operator, and also from higher dimension couplings of the dibaryon
field, that is the dibaryon charge radius. At NLO, we have
contributions coming only from insertions of the kinetic energy in
the deuteron propagator. The NLO corrections to neutron-deuteron
scattering are simply the range corrections in first order
perturbation theory.

At N$^2$LO order there will be corrections to neutron-deuteron
scattering which are simply two times insertions of the kinetic
energy in the deuteron propagator as well as correction from nucleon
structure comes from the nucleon radii\cite{Chen,Beane}. For higher
orders in the EFT expansion, order $Q^3$, there are contributions
from several types of graphs, including relativistic corrections and
contribution from, a four-nucleon-one-photon operator, the dibarion
charge radius operator.

The last diagram also show the photon couples to the three-nucleon
force. At LO calculation, there is no interaction of ${\bf {H_0}}$
with a photon, because $H_0$ has no derivatives, so it is not
affected by the minimal substitution ${\bf P}\rightarrow {\bf P} - e
{\bf A}$. In other hand, contribution from the photon couples to the
three-nucleon field (${\bf {H_2}}$) is calculated at N$^2$LO.

Low-energy observables must be independent of an arbitrary regulator
$\Lambda$ up to the order of the expansion. One can therefore
estimate sensitivity of the results to cut-off $\Lambda$, and hence
provide a reasonable error-analysis, by employing a momentum cut-off
in the solution of the Faddeev equation and varying it between the
breakdown-scale $\LambdaNoPion$ and $\infty$. The cut-off varied
between $\Lambda=200$ MeV and $\Lambda=550$ MeV.

\section{Results and discussion}
\label{section:comparison}

A numerical solution of the charge form factor and charge radius of
triton have been done. The numerical values and parameters used for
calculation is described in ref.~\cite{Sadeghi2}. We solved integral
equation by insertion of $Q$ in Faddeev equation where is shown in
figs.~1 and ~2. The integral equation is solved by the first two
terms of the effective range expansion, properly iterating and
folding to electromagnetic interaction where are shown in fig.~2
order by order and integrated on the involving momentum.  The
three-body parameter ${\bf {H_0}}$ is fixed from the
${}^2\mathrm{S}_\frac{1}{2}$ scattering length
$a_3=(0.65\pm0.04)\;\mathrm{fm}$ at LO. ${\bf {H_2}}$ is also
required at N$^2$LO and it is determined by the triton binding
energy $B_3=8.48$ MeV.

The calculated $^3$H charge form factor is compared with the
experimental data. The results have been shown in fig.~3. The
rhomboids and squares are experimental data from Refs.~\cite{Beck}
and \cite{Beck1}, respectively. While the general trend of the
low-momentum transfer form factor data is reproduced by our
calculation, we obtain a somewhat close slope to the experiments up
to N$^2$LO.
\begin{figure}[!htp]
\begin{center}
\includegraphics*[width=0.6\textwidth]{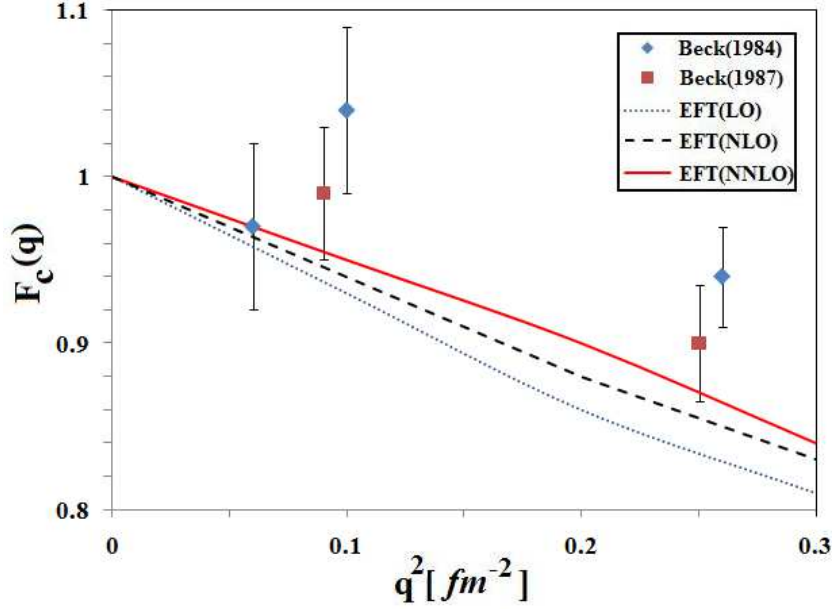} \vspace*{8pt}
\caption{The triton charge form factor. Dot, dash and solid curve
are LO, NLO, and N$^2$LO EFT results. The rhomboids and squares are
experimental data from Refs.~\cite{Beck2} and ~\cite{Beck},
respectively.} \vspace*{-0pt} \label{fig3}
\end{center}
\end{figure}
The remarkable success of the present calculation by using EFT
should be stressed. The number of three-body calculations with
external currents is however, extremely limited. It suggests, in
particular, that the present model provide simple inclusion of
external currents.  Yet, the excellent agreement between the
calculated and measured results of charge form factor suggests that
these corrections may be negligible.

Finally, values for the charge radii of $^3$H are listed in Table
{I}. The results in comparison of the modern nucleon-nucleon
potential AV18 together with the three- nucleon force UrbanaIX, are
found to be in good agreement with experimental data.
\begin{table}[!htb]
\caption{Comparison between different theoretical and experimental
results for triton charge radius in fm.}\vspace{0.25cm}
\begin{center}
 \begin{tabular}{c ||c|c c}
\hline Theory & Triton charge radius & \% Error \\ \hline \hline
Effective theory~\cite{Platter} & 2.1 & 20 \\
Argonne v18+Urbana IX~\cite{Golak} & 1.722  & 2 \\
Argonne v18+Urbana IX~\cite{Marcucci} & 1.725  & 2 \\
EFT(LO) & 1.959  & 12 \\
EFT(NLO) & 1.846 & 5 \\
EFT(N$^2$LO) & 1.775  & 1.1 \\
Exp.~\cite{Amroun}& 1.755$\pm$0.086 & \\ \hline
\end{tabular}
\end{center}
\end{table}
Fitting a polynomial in ${\bf q^2}$ to our result for the charge
form factor, eq.(7), we can extract the triton charge radius.  This
procedure leads to a triton charge radius
$r_c=$[1.959(LO)-0.113(NLO)-0.071(N$^2$LO)]fm =[1.775 $\pm$ 0.02]
fm.

The cut-off variation has been also done. These variations were
nearly independent of variation of momentum and decreased steadily
when the order of calculation is increased up to N$^2$LO.

In fig. 4, we show the correlation between the triton charge radius
and binding energy of triton.  The solid line denotes leading order
effective theory result~\cite{Platter}. The two boxes show modern
calculations (including meson exchange currents) based on the AV18
potential with and without the Urbana IX three-body
force~\cite{Golak}. The circles indicate Faddeev calculations using
different potentials from Refs. ~\cite{Payne,Payne2}.  The
correlation curve of Ref.~\cite{Friar} is given by the dotted line.
The triangle with error-bar indicates the experimental
values~\cite{Amroun}. Short-dash, dot-dash and long-dash curves are
LO, NLO, and N$^2$LO our EFT results, respectively. It should be
considered that the inclusion of the range correction moves the
universal line closer to the experimental values in corresponding to
Phillips line.

\begin{figure}[!htp]
\begin{center}
\includegraphics*[width=0.6\textwidth]{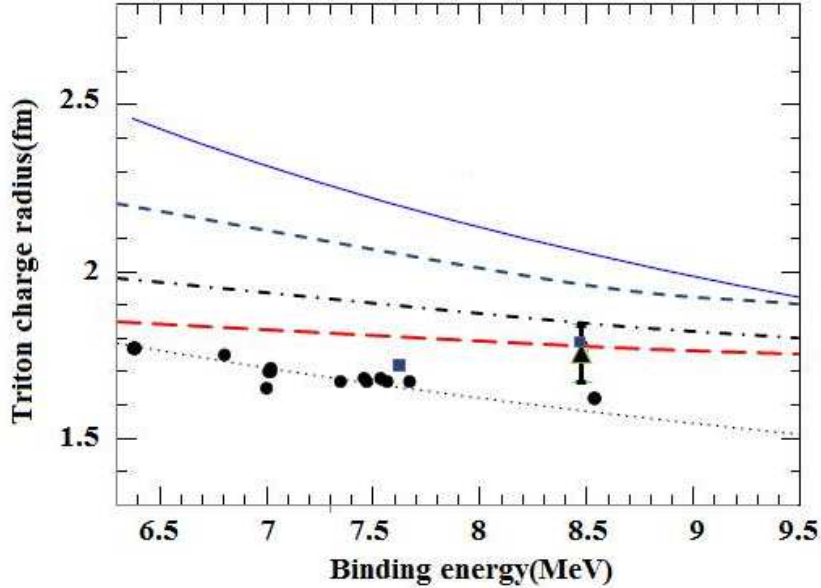} \vspace*{8pt}
\caption{The correlation between the triton charge radius and
binding energy. The solid line denotes  leading order effective
theory result~\cite{Platter}. The circles indicate Faddeev
calculations using different potentials from Refs.
~\cite{Payne,Payne2} while the two boxes show modern calculations
using AV18 with and without the Urbana IX three-body force
~\cite{Golak}. The correlation curve of Ref.~\cite{Friar} is given
by the dotted line. The triangle with error-bar indicates the
experimental values~\cite{Amroun}. Short-dash, dot-dash and
long-dash curves are LO, NLO, and N$^2$LO our EFT results,
respectively. For better comparison, axes scales are chosen like as
~\cite{Platter}. } \vspace*{-0pt} \label{fig4}
\end{center}
\end{figure}
\section{Summery and Conclusion}
\label{section:conclusion}
The goal of this work was to obtain the charge form factor and
charge radius of triton, as a three-body bound state, that are
obtained by using EFT. We applied the formalism developed to the
computation of processes involving external currents such as, triton
charge form factor and charge radii. We considered the diagrams
needed to be computed and the charge form factor numerically
obtained by perturbatively solving of Faddeev equation, up to
N$^2$LO. At very low energies, the interactions between nucleons can
be only described by point-like interactions. The present study of
the triton charge form factor show that it is possible to obtain
remarkably good predictions for electrogenic structure functions of
the few-body bound states by using EFT at very low-energies.

The charge form factor of triton at low energies was calculated by
using pionless EFT and simple inclusion of external currents. The
charge radii is also calculated up to N$^2$LO.  The triton binding
energy and nd scattering length in the triton channel have been used
to fix them. Hence the triton  charge radii is in total determined
as $r_c=$[1.959(LO)-0.113(NLO)-0.071(N$^2$LO)]fm =[1.775 $\pm$ 0.02]
fm. It converges order by order in low energy expansion and is also
cut-off independent at this order. The results also move the
universal line closer to the experimental values when order of
calculation increases. This is in corresponding to the Phillips
line.

\section{Acknowledgments}
The author would like to thank H.W.~Grie{\3}hammer for many helpful
discussions and his available mathematica code. This work is
supported in part by the university of Arak.


\end{document}